\documentclass[pra,aps,aps,showpacs,twocolumn,superscriptaddress]{revtex4}
\usepackage{amsfonts}
\usepackage{amsmath}
\usepackage{amssymb}
\usepackage{graphicx}
\usepackage{txfonts}

\begin{document}

\title{Fisher information of a squeezed-state interferometer with a finite
photon-number resolution}
\author{P. Liu}
\affiliation{Department of Physics, Beijing Jiaotong University, Beijing 100044, China}
\author{P. Wang}
\affiliation{Beijing Computational Science Research Center, Beijing 100084, China}
\author{W. Yang}
\email{wenyang@csrc.ac.cn}
\affiliation{Beijing Computational Science Research Center, Beijing 100084, China}
\author{G. R. Jin}
\email{grjin@bjtu.edu.cn}
\affiliation{Department of Physics, Beijing Jiaotong University, Beijing 100044, China}
\author{C. P. Sun}
\email{cpsun@csrc.ac.cn}
\affiliation{Beijing Computational Science Research Center, Beijing 100084, China}
\date{\today }

\begin{abstract}
Squeezed-state interferometry plays an important role in quantum-enhanced
optical phase estimation, as it allows the estimation precision to be
improved up to the Heisenberg limit by using ideal photon-number-resolving
detectors at the output ports. Here we show that for each individual $N$%
-photon component of the phase-matched coherent $\otimes $ squeezed vacuum
input state, the classical Fisher information always saturates the quantum
Fisher information. Moreover, the total Fisher information is the sum of the
contributions from each individual $N$-photon components, where the largest $%
N$ is limited by the finite number resolution of available photon counters.
Based on this observation, we provide an approximate analytical formula that quantifies the amount of lost information due to the finite photon number resolution, e.g., given the mean photon number $\bar{n}$ in the input state, over $96$
percent of the Heisenberg limit can be achieved with the number resolution
larger than $5\bar{n}$.
\end{abstract}

\maketitle

\section{Introduction}

Quantum-enhanced optical phase estimation through the Mach-Zehnder
interferometer (MZI) is important for multiple areas of scientific research~%
\cite{Caves,Giovannetti,Ma,Aasi,Dowling15,Matthews,Luca}, such as imaging,
sensing, and high-precision gravitational waves detection. The MZI-based
optical phase estimation consists of three steps (see e.g. Fig.~\ref{fig1}%
(a)). First, a two-mode input state of the light is prepared. Second, the
light passes successively through a beam splitter, the unknown relative
phase shift $\varphi$ between the two arms of the MZI, and another beam
splitter, and evolves to the output state. Third, the output state is
measured for many times and the outcomes $\mathbf{x}=\{x_1, x_2, ..., x_v\}$
is processed to construct an unbiased estimator $\hat{\varphi}(\mathbf{x})$
to the unknown parameter $\varphi$~\cite{Helstrom,Kay}. The estimation
precision is quantified by the standard deviation $\Delta \varphi\equiv\sqrt{%
\langle(\hat{\varphi}(\mathbf{x})-\varphi)^{2}\rangle}$. By using optimal
data processing techniques to extract all the information contained in the
data, the estimation precision from $v\gg1$ repeated measurements is given
by the Cram\'{e}r-Rao lower bound~\cite{Helstrom,Kay}: $\Delta\varphi_{%
\mathrm{CRB}}\equiv1/\sqrt{vF(\varphi)}$, where $F(\varphi)$ is the
classical Fisher information (CFI) for the measurement scheme used. Given
the input state, maximizing $F(\varphi)$ over all possible measurement
schemes gives the quantum Fisher information (QFI) $F_{Q}$ and hence the
quantum Cram\'{e}r-Rao bound $\Delta\varphi_{\mathrm{QCRB}}\equiv1/\sqrt{%
vF_{Q}}$~\cite{Braunstein,Braunstein96,Luo,Smerzi09,Donner}, which sets an
ultimate precision for estimating the unknown phase shift $\varphi$.
Usually, the precision $\Delta\varphi_{\mathrm{QCRB}}$ improves with
increasing number of photons $\bar{n}$ contained in the input state. Using a
coherent-state of light as the input, the achievable phase sensitivity per
measurement is limited by the classical (or shot noise) limit $%
\delta\varphi\equiv\sqrt{v}\Delta \varphi \sim 1/\sqrt{\bar{n}}$, as the QFI
$F_{Q}\sim O(\bar{n})$.

To improve the precision beyond the classical limit ($\sim1/\sqrt{\bar{n}}$%
), it is necessary to employ quantum resources, such as entanglement and
squeezing in the input state~\cite%
{Caves,Giovannetti,Ma,Aasi,Dowling15,Matthews,Luca}. In this context, the
squeezed states of light play an important role and have been widely studied
in the past decades ever since the pioneer work of Caves in 1981~\cite{Caves}%
, who shows that by feeding a coherent state $|\alpha\rangle$ into one port
of the MZI and a squeezed vacuum $|\xi\rangle$ into the other port, the
unknown phase shift can be estimated with a precision beyond the classical
limit. In 2008, Pezz\'{e} and Smerzi~\cite{Smerzi08} further suggested that
the previously used phase estimator based on the averaged relative photon
number is not optimal. When the injected fields are phase matched, i.e., the
phases of two light fields $\theta_a$ and $\theta_b$ obeying $%
\cos(\theta_{b}-2\theta_{a})=+1$, the QFI can reach the Heisenberg scaling $%
\sim O(\bar{n}^{2})$ for a given mean photon number $\bar{n}%
=|\alpha|^{2}+\sinh^{2}|\xi|$. More importantly, this QFI can be saturated
by the CFI for ideal photon counting measurements. Consequently, by using
the optimal data processing technique (such as the maximum-likelihood
estimation or Bayesian estimation) to process these measurement outcomes,
the phase estimation precision can attain the Heisenberg limit $%
\delta\varphi_{\mathrm{CRB}}=\delta \varphi_{\mathrm{QCRB}}\sim1/\bar{n}$.
Recently, Lang and Caves~\cite{Lang} proved that given the total average
photon number $\bar{n}$ of the input state, if a coherent light is fed into
one input port of the MZI, then the squeezed vacuum is the optimal state to
inject into the second input port. Liu \textit{et al}.~\cite{Liu} have
analyzed the phase-matching condition (PMC) that maximizes the QFI in the
squeezed-state interferometer, where a superposition of even or odd number
of photons is injected from one port of the interferometer and any input
state from another.

An important requirement of these theoretical works~\cite{Smerzi08,Lang} is
to take into account all the photon-counting events, which in turn requires
photon-number-resolving detectors with perfect number resolution~\cite%
{Seshadreesan}. However, on the experimental side, the best detector up to
date can only resolve the number of photons up to $4$~\cite%
{Smerzi07,Kardynal}. This makes it unclear whether or not the Heisenberg
limit of the estimation precision can still be achieved by using realistic
photon detectors with an upper threshold on the number resolution. To bridge
this gap between the theory and experiments, it is of interest to
investigate the experimentally achievable estimation precision when the
total number of photons being detected is limited, i.e., $N=N_{1}+N_{2}\leq
N_{\mathrm{res}}$, where $N_{\mathrm{res}}/2$ determines the number
resolution by a single photon-counting detector. Since the existence of an
upper threshold $N_{\mathrm{res}}$ essentially amounts to discarding the
information contained in photon-counting events with the number of photons
larger than $N_{\mathrm{res}}$, it is therefore important to investigate the
distribution of the QFI and CFI in the $N$-photon components of the coherent
$\otimes$ squeezed vacuum input state and calculate how much the QFI is kept
with a finite number resolution.

In addition, studying the distribution of the QFI and CFI in the $N$-photon
components also helps to understand the phase estimation precision in recent
post-selection experiments. When the MZI is fed by the coherent $\otimes$
squeezed vacuum, the state after the first beam splitter of the MZI contains
a small fraction of the path-entangled NOON state~\cite{Hofmann,Afek}, which
is a well-known $N$-photon non-classical state that allows the phase
estimation precision to achieve the Heisenberg limit~\cite%
{Kok,Pryde,Boto,Mitchell,Walther,Cable,Toppel}. In the limit $%
|\alpha|^{2},|\xi|\ll1$, Afek~\textit{et al.}~\cite{Afek} have demonstrated $%
N$-fold oscillations of the coincidence rates for $N$ up to $5$, manifesting
the appearance of $N$-photon NOON states. However, the generation
probability of a $N$-photon NOON state decreases dramatically with
increasing $N$, e.g., the $5$-photon count rate $\sim3$ per $100$ second~%
\cite{Afek}. Therefore, it is desirable to study the overall estimation
precision when such small generation probabilities are included, since there
are general conclusions that the generated state under postselection cannot
improve the precision for estimating a single parameter when the \textit{%
total} number of input photons are included (see e.g., Refs.~\cite%
{Combes,Pang,Haine}).

In this paper, we investigate the distribution of the QFI and CFI in the
different $N$-photon components of the coherent $\otimes $ squeezed vacuum
input state and provide the achievable estimation precision by using
imperfect photon counters with an upper threshold $N_{\mathrm{res}}$ for the
photon number resolution. Under the PMC $\cos (\theta _{b}-2\theta _{a})=+1$%
, we show that the CFI always saturates the QFI for each \textit{individual}
$N$-photon component. Consequently, when the detectable number of photons is
upper bounded by $N_{\mathrm{res}}$, the phase estimation precision $\delta
\varphi _{\mathrm{CRB}}$ is always equal to $\delta \varphi _{\mathrm{QCRB}}$
and both of them are determined by the sum of the CFI or equivalently the
QFI for each $N$-photon component with $N$ up to $N_{\mathrm{res}}$. For the
commonly used optimal input state with $|\alpha |^{2}\simeq \sinh ^{2}|\xi
|\simeq \bar{n}/2$~\cite{Smerzi08,Lang,Liu}, photon counting measurement
with ideal photon detectors ($N_{\mathrm{res}}\rightarrow \infty $) gives
the CFI or the QFI $F_{Q,\mathrm{opt}}^{\mathrm{(id)}}\sim \bar{n}^{2}$,
leading to the Heisenberg limit of the estimation precision~\cite%
{Smerzi08,Lang,Liu}. For finite photon number resolution, we provide an approximate analytical expression that quantifies
the amount of lost information, which predicts that over $96$ percent of the ideal QFI can be achieved as long as $N_{\mathrm{res}}\gtrsim 5\bar{n}$.
Compared with the ideal case (i.e., $|\alpha
|^{2}\simeq \sinh ^{2}|\xi |$), we find that the optimal input state
contains more coherent light photons than that of the squeezed light.

\section{Finite $N$-photon state under postselection}

\begin{figure}[hptb]
\begin{centering}
\includegraphics[width=1\columnwidth]{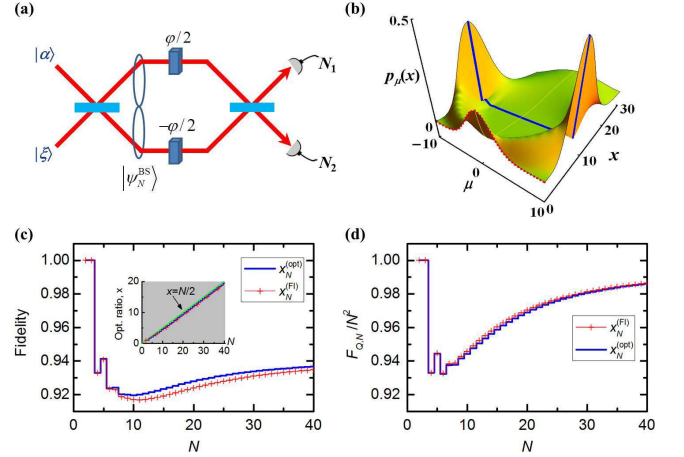}
\caption{(a) Photon counting measurement at output ports of the MZI that fed with a coherent state $|\protect\alpha%
\rangle$ and a squeezed vacuum $|\xi\rangle$, and the $N$-photon state $|\psi_{N}^{\mathrm{BS}}\rangle$, post-selected by the number of photons being detected $N=N_1+N_2=2J$.
(b) For a given $J=N/2=10$, the probability distribution $p_{\mu}=|\langle J, \mu|\psi _{N}^{\mathrm{BS}}\rangle|^2$ against $\mu%
=(N_1-N_2)/2\in[-J, +J]$ and $x=|\protect\alpha|^2/\tanh|\protect\xi|$. At a certain value of the ratio $x_N^{(\mathrm{opt})}$ (indicated by the blue line), the distribution shows two symmetric peaks at $\mu=\pm J$, indicating the appearance of a path-entangled NOON state. (c) The fidelity between the $N$-photon state and an ideal NOON state, and (d) the QFI of $N$-photon state, with their values calculated at $x=x_N^{(\mathrm{opt})}$ (blue solid line) and $x_N^{(\mathrm{FI})}$ (red line with crosses), see text. The inset in (c) indicates $x_N^{(\mathrm{FI})}\leq x_N^{(\mathrm{opt})}\leq N/2$.}\label{fig1}
\end{centering}
\end{figure}

As illustrated schematically by Fig.~\ref{fig1}(a), a post-selection scheme
for creating path-entangled NOON states has been proposed by injecting a
coherent state of light and a squeezed vacuum into Mach-Zehnder
interferometer~\cite{Hofmann,Afek}. This scheme has been demonstrated by
Afek~\textit{et al.}~\cite{Afek} in the limit $|\alpha |^{2},|\xi |\ll 1$.
However, the generated $N$-photon state in postselection cannot improve the
precision for estimating an unknown phase shift, since the CFI is weighted
by the generation probability~\cite{Combes}. It is therefore important to
investigate whether or not a sum of each $N$-component for $N$ up to a
finite number can beat the shot-noise scaling $\sim O(\bar{n})$. To answer
this question, in this section, we first derive explicit form of the $N$%
-photon state generated by postselection. Next, we calculate (quantum)
Fisher information of the $N$-photon state, which determines the ultimate
precision on the phase estimation.

\subsection{The fidelity of the $N$-photon state and the NOON state}

Without any loss and additional reference beams, the input state can be
expressed as a superposition of $N$-photon states~\cite{Donner}, i.e., $%
|\alpha \rangle _{a}\otimes |\xi \rangle _{b}=\sum_{N}\sqrt{G_{N}}|\psi
_{N}\rangle $, where $G_{N}$ denotes the generation probability of a finite $%
N$-photon state, and $N=N_{1}+N_{2}$ is the number of photons post-selected
by the photon counting events $\{N_{1},N_{2}\}$. In Fock basis, the $N$%
-photon state is given by
\begin{equation}
\left\vert \psi _{N}\right\rangle =\frac{1}{\sqrt{G_{N}}}%
\sum_{k=0}^{[N/2]}c_{N-2k}(\theta _{a})s_{2k}(\theta _{b})|N-2k,2k\rangle
_{a,b},  \label{psi_N}
\end{equation}%
where $|m,n\rangle _{a,b}\equiv |m\rangle _{a}\otimes |n\rangle _{b}$, and
the sum over $k$ is up to $[N/2]=(N-1)/2$ (for odd $N$), or $N/2$ (for even $%
N$), because of even number of photons that injected from the port $b$. Note
that the probability amplitudes of the coherent state and the squeezed
vacuum $c_{m}(\theta _{a})=\langle m|\alpha \rangle $ and $s_{n}(\theta
_{b})=\langle n|\xi \rangle $ depend explicitly on the phases of two input
light fields $\theta _{a}$ and $\theta _{b}$ (see the Append. A).
Furthermore, the generation probability $G_{N}$ is also the normalization
factor and is given by
\begin{equation}
G_{N}=\sum_{k=0}^{[N/2]}\left\vert c_{N-2k}s_{2k}\right\vert ^{2}=\frac{%
e^{-\left\vert \alpha \right\vert ^{2}}}{\cosh |\xi |}\left( \frac{\tanh
|\xi |}{2}\right) ^{N}R_{N}(x),  \label{GN}
\end{equation}%
where we have introduced a ratio $x\equiv |\alpha |^{2}/\tanh |\xi |$, and a
polynomial
\begin{equation}
R_{N}(x)=\sum_{k=0}^{[N/2]}\frac{(2k)!}{\left( N-2k\right) !(k!)^{2}}%
(2x)^{N-2k}.  \label{RN}
\end{equation}%
which obeys $R_{N}(0)=N!/[(N/2)!]^{2}$ for even $N$, and $R_{N}(0)=0$ for
odd $N$, similar to the Hermite polynomials at $x=0$. In the limit $|\alpha
|^{2},|\xi |\ll 1$, the ratio can be approximated as $x\sim |\alpha
|^{2}/|\xi |$, and its square is indeed the two-photon probability of the
coherent state divided by that of the squeezed vacuum~\cite{Afek}.

Explicit form of the $N$-photon state crucially depends on the relative
phase difference between the squeezing parameter $\xi$ and the
coherent-state amplitude $\alpha$. Following Refs.~\cite{Smerzi08,Liu}, we
consider the PMC, i.e., $\cos (\theta _{b}-2\theta _{a})=+1$, for which Eq.~(%
\ref{psi_N}) can be reexpressed as $|\psi _{N}\rangle =\exp (iN\theta _{a})|%
\tilde{\psi}_{N}\rangle $, where $|\tilde{\psi}_{N}\rangle $ denotes the $N$%
-photon states with real amplitudes (for details, please see the Append. A).
After the first beam splitter, the $N$-photon state becomes
\begin{equation}
|\psi _{N}^{\mathrm{BS}}\rangle =e^{-i\pi J_{x}/2}|\psi _{N}\rangle
=\sum_{\mu =-J}^{+J}\langle J,\mu \vert \psi _{N}^{\mathrm{BS}}\rangle
|J,\mu \rangle ,  \label{psiBS}
\end{equation}%
where, for brevity, we have introduced the eigenstates of $J_{z}$, i.e., $%
|J,\mu \rangle \equiv |J+\mu ,J-\mu \rangle _{a,b}$, with $J=N/2$ and $\mu
\in \lbrack -J,+J]$. Under the PMC, the probability amplitudes of $|\psi
_{N}^{\mathrm{BS}}\rangle $ can be written as
\begin{equation}
\langle J,\mu \vert \psi _{N}^{\mathrm{BS}}\rangle =e^{iN\theta _{a}}e^{i\pi
(\mu -J)/2}\sqrt{p_{\mu }},  \label{ampBS}
\end{equation}%
which depends solely on the phase of the coherent-state light $\theta _{a}$,
and the probability distribution (see the Append. A)
\begin{widetext}
\begin{eqnarray}
p_{\mu }\equiv |\langle J,\mu |\psi _{N}^{\mathrm{BS}}\rangle |^{2} =\frac{1%
}{R_{N}(x)}\left( \sum_{k=0}^{[N/2]}d_{\mu ,J-2k}^{J}\left( \frac{\pi }{2}%
\right) \frac{\sqrt{(2k)!}}{k!\sqrt{(N-2k)!}}\left( 2x\right)
^{N/2-k}\right) ^{2},  \label{pmu}
\end{eqnarray}
\end{widetext}
where $d_{\mu ,v}^{J}(\varphi )$ are the elements of Wigner's d-matrix~\cite%
{Tajima,Feng}. It is interesting to note that for a given $N$, the
probability distribution depends only on the introduced ratio $x=|\alpha
|^{2}/\tanh |\xi |$; hereinafter, denoted by $p_{\mu }=p_{\mu }(x)$.

Figure~\ref{fig1}(b) shows the probability distribution as a function of $%
\mu $ for a large enough $N$. At $x=0$, i.e., a pure squeezed vacuum being
injected, the probability distribution is almost a Gaussian, due to $p_{\mu
}(0)=[d_{\mu ,-J}^{J}(\pi /2)]^{2}\varpropto \exp (-\mu ^{2}/J)$. As $x$
increases, the $N$-photon state always shows symmetric probability distribution
(i.e., $p_{-\mu }=p_{+\mu }$). One can see this directly from Eq.~(\ref{pmu}%
), where $d_{-\mu ,v}^{J}(\varphi )=(-1)^{J-v}d_{+\mu ,v}^{J}(\pi -\varphi )$%
; see e.g., Refs.~\cite{Tajima,Feng}. Physically, the symmetric probability
distribution arises from the fact that the $N$-photon state $|\psi
_{N}\rangle $ contains only even number of photons in mode $b$, i.e., $%
\langle \psi _{N}|J_{y}|\psi _{N}\rangle =\mathrm{Im}\langle \psi
_{N}|a^{\dagger }b|\psi _{N}\rangle =0$, which in turn leads to
\begin{equation}
\langle \psi _{N}|J_{y}|\psi _{N}\rangle =\langle \psi _{N}^{\mathrm{BS}%
}|J_{z}|\psi _{N}^{\mathrm{BS}}\rangle =\sum_{\mu \geq 0}(p_{+\mu }-p_{-\mu
})\mu =0,  \label{zero}
\end{equation}%
and hence $p_{-\mu }=p_{+\mu }$. This symmetry enables us to write down
explicit expression of the $N$-photon state 
\begin{equation}
|\psi _{N}^{\mathrm{BS}}\rangle \!=e^{iN\theta _{a}}\sum_{\mu \geq 0}e^{i%
\frac{\pi }{2}(\mu -J)}\sqrt{2p_{\mu }(x)}\left( \frac{|J,\mu \rangle
+e^{-i\pi \mu }|J,-\mu \rangle }{\sqrt{2}}\right) ,  \label{psi_BS}
\end{equation}%
which is indeed a superposition of the path-entangled states $\sim (|J,\mu
\rangle +e^{-i\pi \mu }|J,-\mu \rangle )$, where the relative phase $%
e^{-i\pi \mu }$ comes from Eq.~(\ref{ampBS}). For a certain value of $x$,
the probability distribution $p_{\mu }(x)$ reaches its maximum at $\mu =\pm
J=\pm N/2$, indicating $|\psi _{N}^{\mathrm{BS}}\rangle \rightarrow |\psi _{%
\mathrm{NOON}}\rangle =(|J,+J\rangle +e^{-i\pi J }|J,-J\rangle )/\sqrt{2}$,
with the fidelity given by
\begin{equation}
\mathcal{F}_{\mathrm{NOON}}\equiv \left|\langle \psi _{\mathrm{NOON}}|\psi
_{N}^{\mathrm{BS}}\rangle \right|^{2}=2p_{J}(x).  \label{fidelity}
\end{equation}%
Clearly, the fidelity depends on the ratio $x$ and the number of photons
being detected $N$ ($=2J$). For a given $N$, maximizing the fidelity with
respect to $x$, one can obtain the optimal value of the ratio, denoted
hereinafter as $x_{N}^{\mathrm{(opt)}}$. For small $N$'s, it has been
obtained $x_{N}^{\mathrm{(opt)}}=1$ (for $N=2$, $3$), $\sqrt{3}$ ($N=4$),
and $2.016$ ($N=5$); see Ref.~\cite{Afek}. When $N\gg 1$, the optimal value
of $x$ is about $N/2$, for which $\mathcal{F}_{\mathrm{NOON}}\rightarrow
\sqrt{8/9}\simeq 0.943$ (see Ref.~\cite%
{Hofmann}, and also Table~\ref{optimal}). In Fig.~\ref{fig1}(c), we show the optimal value of the fidelity $\mathcal{F}_{%
\mathrm{NOON}}(x_{N}^{\mathrm{(opt)}})$ as a function of $N$ (the blue solid
line), which coincides with Afek \textit{et al}~\cite{Afek}. From Eqs.~(\ref%
{psiBS}) and (\ref{fidelity}), one can also see that before the first beam
splitter, $|\psi _{N}\rangle $ itself at $x=x_{N}^{\mathrm{(opt)}}$
approaches the NOON state $\exp (i\pi J_{x}/2)|\psi _{\mathrm{NOON}}\rangle $%
, which shows the polarization along $\pm J_{y}$.

\subsection{The Fisher information of the post-selected $N$-photon state}

We now investigate the CFI of the $N$-photon state in the photon-counting
measurements and show that it always equals to the QFI under the PMC. To
this end, we first calculate the QFI of the phase-encoded state $\exp
(-i\varphi J_{y})|\psi _{N}\rangle $, where the unitary operator represents
sequent actions of the first beam splitter, the phase-shift accumulation in
the path, and the second 50:50 beam splitter at the output ports, as illustrated
in Fig.~\ref{fig1}(a). Due to $\langle \psi _{N}|J_{y}|\psi _{N}\rangle =0$,
it is easy to obtain the QFI~\cite%
{Braunstein,Braunstein96,Luo,Smerzi09,Donner}:
\begin{equation}
F_{Q,N}=4\langle \psi _{N}|J_{y}^{2}|\psi _{N}\rangle =4\langle \psi _{N}^{%
\mathrm{BS}}|J_{z}^{2}|\psi _{N}^{\mathrm{BS}}\rangle =4\sum_{\mu
=-J}^{+J}\mu ^{2}p_{\mu },  \label{FQN}
\end{equation}%
where $|\psi _{N}^{\mathrm{BS}}\rangle $ denotes the $N$-photon state after
the first beam splitter and its probability distribution $p_{\mu }(x)$ has
been given by Eq.~(\ref{pmu}). Similar to the fidelity, one can see that the
QFI depends on the ratio $x$ and the number of photons $N$. For the cases $%
N=2$, $3$, and $4$, both of them reach maximum at $x_{N}^{(\mathrm{opt})}$.
This is because of the relation:
\begin{equation}
F_{Q,N}=N^{2}\mathcal{F}_{\mathrm{NOON}}(x)+2(N-2)^{2}p_{J-1}(x)+\cdots ,
\end{equation}%
where $p_{0}(x)$, $p_{1/2}(x)$, and $p_{1}(x)$ are vanishing at $x=x_{N}^{(%
\mathrm{opt})}$. When $N$ $\geq 5$, however, $\{p_{|\mu |}(x)\}$ with $|\mu
|<J$ provide nonvanishing contributions to the QFI. Numerically, we find
that $F_{Q,N}$ reaches its maximum at $x_{N}^{(\mathrm{FI})}$, which is
slightly smaller than $x_{N}^{(\mathrm{opt})}$ (see Table I). In Fig.~\ref%
{fig1}(d), we plot maximum of the QFI as a function of $N$ and find $F_{Q,N}$
$\gtrsim 0.933N^{2}$.

Next, we consider the photon counting measurements over the phase-encoded state $\exp (-i\varphi
J_{y})|\psi _{N}\rangle$ and calculate the CFI. Again, we consider the PMC and rewrite the $N$%
-photon state as $|\psi _{N}\rangle =\exp (iN\theta _{a})|\tilde{\psi}%
_{N}\rangle $, where $|\tilde{\psi}_{N}\rangle $ is given by Eq.~(\ref{psi_N}%
) with $\theta _{a}=\theta _{b}=0$. Note that the probability amplitudes of $%
|\tilde{\psi}_{N}\rangle$ and hence that of $\exp (-i\varphi J_{y})|\tilde{%
\psi}_{N}\rangle $ are real, which result in the conditional probabilities
(see the Append. A):
\begin{equation}
P_{N}(\mu |\varphi )=|\langle J,\mu |e^{-i\varphi J_{y}}|\psi _{N}\rangle
|^{2}=\left[ \langle J,\mu |e^{-i\varphi J_{y}}|\tilde{\psi}_{N}\rangle %
\right] ^{2},  \label{PNmu}
\end{equation}%
where $\mu =(N_{1}-N_{2})/2\in \lbrack -J,+J]$ and $J=(N_{1}+N_{2})/2=N/2$.
Obviously, for a given $N$, there are $N+1$ outcomes with their probabilities satisfying the normalization condition $\sum_{\mu }P_{N}(\mu
|\varphi )=\langle \psi _{N}|\psi _{N}\rangle =1$. Due to the real
probability amplitudes, i.e., $\langle J,\mu |\exp (-i\varphi J_{y})|\tilde{%
\psi}_{N}\rangle \in \mathbb{R}$, we further obtain
\begin{widetext}
\begin{equation*}
\frac{\partial P_{N}(\mu |\varphi )}{\partial \varphi }=2\sqrt{P_{N}(\mu
|\varphi )}\langle J,\mu |(-iJ_{y})\exp (-i\varphi J_{y})|\tilde{\psi}%
_{N}\rangle \in \mathbb{R},
\end{equation*}%
indicating that $\langle J,\mu |J_{y}\exp (-i\varphi J_{y})|\tilde{\psi}%
_{N}\rangle $ is purely imaginary for each $\mu $. This is the key point to
obtain the CFI:
\begin{equation}
F_{N}(\varphi )=\sum_{\mu =-J}^{+J}\frac{\left[ \partial P_{N}(\mu |\varphi
)/\partial \varphi \right] ^{2}}{P_{N}(\mu |\varphi )}=-4\sum_{\mu =-J}^{+J}%
\left[ \langle J,\mu |J_{y}e^{-i\varphi J_{y}}|\tilde{\psi}_{N}\rangle %
\right] ^{2}=4\langle \tilde{\psi}_{N}|J_{y}^{2}|\tilde{\psi}_{N}\rangle
=F_{Q,N},  \label{FQN2}
\end{equation}%
where $F_{Q,N}$ is the QFI of the phase-encoded state $\exp (-i\varphi
J_{y})|\psi _{N}\rangle $ under the PMC, given by Eq.~(\ref{FQN}).

\begin{table}[hptb]
\caption{For a given $N$, the fidelity $\mathcal{F}_{\mathrm{NOON}}$ and the QFI $F_{Q,N}$ depend solely on the ratio $x\equiv |\protect%
\alpha |^{2}/\tanh|\protect\xi|$, and reach maximum at $x_{N}^{(\mathrm{opt}%
)}$ and $x_{N}^{(\mathrm{FI})}$, respectively. For $N=2$, $3$, $\mathcal{F}_{\mathrm{%
NOON}}=F_{Q,N}/N^{2}=1$ at $x_{N}^{(\mathrm{opt})}=x_{N}^{(\mathrm{FI})}=1$;
While for $N=4$, $\mathcal{F}_{\mathrm{NOON}}=F_{Q,N}/N^{2}=0.933$ at $x_{N}^{(\mathrm{%
opt})}=x_{N}^{(\mathrm{FI})}=\protect\sqrt{3}$.}
\label{optimal}\vspace{0.2cm}
\begin{tabular}{ccccccccccccccccc}
\hline\hline
$N$ &  & 5 &  & 6 &  & 7 &  & 8 &  & 9 &  & 10  &  & 100 \\ \hline
$x_{N}^{(\mathrm{opt})}$, $x_{N}^{(\mathrm{FI})}$ &  & 2.016, 1.962 &  &
2.544, 2.488 &  & 2.961, 2.856 &  & 3.444, 3.323 &  & 3.908, 3.752 &  &
4.390, 4.213  &  & 49.405, 49.103 \\[0.1cm]
$\mathcal{F}_{\mathrm{NOON}}$, $F_{Q,N}/N^2$ &  & 0.941, 0.945 &  & 0.924, 0.933 &  &
0.924, 0.938 &  & 0.920, 0.939 &  & 0.920, 0.943 &  & 0.920, 0.946  &  & 0.941, 0.995 \\[0.1cm] \hline\hline
\end{tabular}
\vspace{0.2cm}
\end{table}
\end{widetext}

As one of main results of this work, Eq.~(\ref{FQN2}) indicates that as the
``input" state, $|\psi _{N}\rangle$ at $x=x_{N}^{(\mathrm{FI})}$ could
provide a global phase estimation at the Heisenberg scaling~\cite{Hofmann09}%
, as $F_{N}(\varphi )=F_{Q,N}\gtrsim 0.933N^{2}$. However, this scaling is
defined with respect to the number of photons being detected $N$.
Furthermore, $|\psi _{N}\rangle$ is post-selected by the $N$-photon
detection events with the generation probability $G_{N}$, which is usually
very small as $N\gg 1$ (see Fig.~\ref{fig2}(a)). Indeed, purely with the $N$%
-photon detection events (i.e., totally $N+1$ outcomes with a definite $N$),
one cannot improve the accuracy for estimating an unknown phase shift, since
the CFI is weighted by the generation probability~\cite{Combes}, i.e., $%
G_{N}F_{N}(\varphi)$. For the input $|\alpha \rangle _{a}\otimes |\xi
\rangle _{b}$ with a given mean photon number $\bar{n}=|\alpha |^{2}+\sinh
^{2}|\xi|$, one can see that the weighted CFI for different values of $N$
can only reach the classical limit $\sim O(\bar{n})$, as depicted by Fig.~%
\ref{fig2}(c), where we considered the special case $\alpha, \xi\in \mathbb{R}$, for which the PMC is naturally fulfilled and therefore $F_{N}(\varphi)=F_{Q,N}$.

\section{The total Fisher information}

In order to improve the estimation precision, all the detection evens $\{N_{1},N_{2}\}$
have to be taken into account in the photon-counting measurements, which
gives ideal result of the CFI~\cite%
{Braunstein,Braunstein96,Luo,Smerzi09,Donner}:
\begin{equation}
F^{\mathrm{(id)}}(\varphi )=\sum_{2J=0}^{\infty }\sum_{\mu =-J}^{+J}\frac{%
\left[ \partial P(J,\mu |\varphi )/\partial \varphi \right] ^{2}}{P(J,\mu
|\varphi )}=\sum_{N=0}^{\infty }G_{N}F_{N}(\varphi ),  \label{CFI_CV}
\end{equation}%
where we have reexpressed the input state as $|\psi _{\mathrm{in}}\rangle
=\sum_{N}\sqrt{G_{N}}|\psi _{N}\rangle $, so we have
\begin{equation*}
P(J,\mu |\varphi )\equiv |\langle J,\mu |e^{-i\varphi J_{y}}|\psi _{\mathrm{%
in}}\rangle |^{2}=G_{N}P_{N}(\mu |\varphi ),
\end{equation*}%
and $P_{N}(\mu |\varphi )\equiv |\langle J,\mu |\exp (-i\varphi J_{y})|\psi
_{N}\rangle |^{2}$, given by Eq.~(\ref{PNmu}). Note that the total CFI is
indeed a sum of each $N$-component contribution $F_{N}(\varphi )$ weighted
by $G_{N}$. With only the $N$-photon detection events, the Fisher
information is simply given by $G_{N}F_{N}(\varphi )$, as mentioned above.

\begin{figure}[hptb]
\begin{centering}
\includegraphics[width=1\columnwidth]{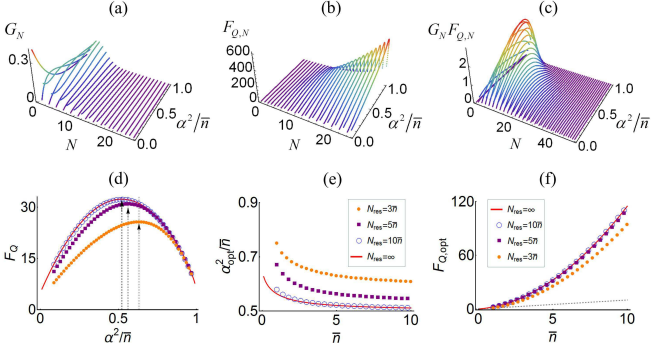}
\caption{(a) The generation probability $G_N$, (b) the QFI of each $N$-photon state $F_{Q,N}$, (c) the weighted QFI $G_N F_{Q,N}$, and (d) the total QFI $F_{Q}(N_{\mathrm{res}}, \bar{n}, \alpha^{2})$, for $\bar{n}=5$ fixed and $N_{\mathrm{res}}=3\bar{n}$ (solid circles), $5\bar{n}$ (squares), $10\bar{n}$ (open circles), and $\infty$ (red solid lines). The last case is given by Eq.~(\ref{QFI}), which predicts $\alpha_{\mathrm{opt}}^{2}\simeq \bar{n}/2$ and $F_{Q,\mathrm{opt}}^{\mathrm{(id)}}\simeq \bar{n}(\bar{n}+3/2)$. For each a given $\bar{n}\in [1, 10]$, using the same values of $N_{\mathrm{res}}$ and maximizing the total QFI with respect to $\alpha^2$ to obtain: (e) $\alpha^2_{\mathrm{opt}}/\bar{n}$, and (f) the associated QFI $F_{Q,\mathrm{opt}}$. In (c), the peak height of the weighted QFI is about $\bar{n}/2$. The vertical lines in (d): the optimal value of $\alpha^2$ for different values of $N_{\mathrm{res}}$. The dashed line in (f): the classical limit $F_Q=\bar{n}$.} \label{fig2}
\end{centering}
\end{figure}

Similar to Eq.~(\ref{FQN}), we further calculate the total QFI of the output
state $\exp(-i\varphi J_{y})|\psi _{\mathrm{in}}\rangle$, which is
independent from any specific measurement scheme and is given by $%
F_{Q}=4(\langle J_{y}^{2}\rangle _{\mathrm{in}}-\langle J_{y}\rangle_{%
\mathrm{in}}^{2})$~\cite{Braunstein,Braunstein96,Luo,Smerzi09,Donner}. For
the input state $|\alpha \rangle _{a}\otimes |\xi \rangle _{b}$, we obtain $%
\langle J_{y}\rangle_{\mathrm{in}}=0$, and hence ideal result of the QFI
\begin{equation}
F_{Q}^{\mathrm{(id)}}=4\sum_{N=0}^{\infty }G_{N}\langle \psi
_{N}|J_{y}^{2}|\psi _{N}\rangle =\sum_{N=0}^{\infty }G_{N}F_{Q,N},
\label{FI_CV}
\end{equation}%
where $F_{Q,N}$ is the QFI of the $N$-photon component. Under the PMC, we
have show that for each $N$-component $F_{N}(\varphi )=F_{Q,N}$, which
naturally results in a global phase estimation $F^{\mathrm{(id)}%
}(\varphi)=F_{Q}^{\mathrm{(id)}}$~\cite{Hofmann09}. According to Refs.~\cite%
{Smerzi08,Lang,Liu}, one can obtain explicit form of the QFI by directly
calculating $4\langle J_{y}^{2}\rangle _{\mathrm{in}}$ (see also the Append.
B), namely
\begin{equation}
F^{\mathrm{(id)}}(\varphi )=F_{Q}^{\mathrm{(id)}}=|\alpha |^{2}e^{2|\xi
|}+\sinh ^{2}|\xi |.  \label{QFI}
\end{equation}%
Given a constraint on the mean photon number $\bar{n}$, maximum of the QFI
was found to achieve the Heisenberg scaling $F_{Q,\mathrm{opt}}^{\mathrm{(id)%
}}\simeq \bar{n}(\bar{n}+3/2)\sim O(\bar{n}^{2})$~\cite{Lang}, provided $%
|\alpha |^{2}\simeq\sinh ^{2}|\xi |\simeq\bar{n}/2\gg 1$~\cite{Smerzi08};
see also the red solid lines of Fig.~\ref{fig2}(d)-(f). However, such a
scaling is only possible with exactly perfect photon-number-resolving
detectors~\cite{Seshadreesan}, which enable us to record infinite number of
the photon-counting events; see also Eq.~(\ref{CFI_CV}).

Usually, a single number-resolving detector can only register the number of
photons up to $4$~\cite{Smerzi07,Kardynal}. It is therefore important to
investigate the CFI of each $N$-component for $N$ up to a finite number of
photons being resolvable $N_{\mathrm{res}}$. For brevity, we consider the
input fields with the real amplitudes and large enough mean photon number
(i.e., $\bar{n}=\alpha ^{2}+\sinh ^{2}\xi >1$). Since the PMC is naturally
fulfilled, the CFI is still a sum of each $N$-component with the weight $%
G_{N}$ and equals to the QFI:
\begin{widetext}
\begin{eqnarray}
F_{Q} =4\sum_{N=0}^{N_{\mathrm{res}}}G_{N}\langle \psi _{N}|J_{y}^{2}|\psi
_{N}\rangle =\sum_{N=0}^{N_{\mathrm{res}}}G_{N}F_{Q,N} =\sum_{N=0}^{N_{\mathrm{res}}}\sum_{k=0}^{[N/2]}\left[ N+4k(N-2k)+\frac{%
4k\alpha ^{2}}{\tanh \xi }\right] \left[ c_{N-2k}(0)s_{2k}(0)\right] ^{2},
\label{QFIres}
\end{eqnarray}%
\end{widetext}
where $|\psi _{N}\rangle$ is the $N$-photon state and $G_{N}=G_{N}(\alpha ^{2},\xi )$ denotes its generation probability, given by
Eqs.~(\ref{psi_N}) and (\ref{GN}). Obviously, the QFI considered
here depends on three variables $\{N_{\mathrm{res}},\alpha ^{2},\xi \}$, or
equivalently, $\{N_{\mathrm{res}},\alpha ^{2},\bar{n}\}$ for a given $\bar{n}
$. When $N_{\mathrm{res}}\rightarrow \infty $, the ideal result of the QFI
is recovered (see the Append. B).

The Heisenberg scaling of the QFI can be maintained for large enough $N_{%
\mathrm{res}}$, provided that all the nonvanishing $\{G_{N}F_{Q,N}\}$ are
included. To obtain the minimum value of $N_{\mathrm{res}}$, we show $G_{N}$%
, $F_{Q,N}$, and $G_{N}F_{Q,N}$ against $N$ and $\alpha ^{2}$ under a
constraint on $\bar{n}$. From Fig.~\ref{fig2}(b), one can see that $F_{Q,N}$
increases quadratically with $N$. This is because the QFI reaches its
maximum $F_{Q,N}\sim O(N^{2})$ when $\alpha ^{2}/\tanh \xi =x_{N}^{(\mathrm{%
FI})}$ (see Table I), which corresponds to $\alpha ^{2}/\bar{n}\rightarrow 1$%
, i.e., the classical light being dominant for a given $\bar{n}=\alpha
^{2}+\sinh ^{2}\xi $. On the other hand, the generation probability shows a
little complex behavior on $N$; see Fig.~\ref{fig2}(a). At $\alpha ^{2}=0$, $%
G_{N}$ is nonvanishing at even number of $N$ and decreases monotonically
with the increase of $N$. When $\alpha ^{2}\geq 1$ (i.e., $G_{1}\geq G_{0}$%
), it reaches maximum at a certain value of $N$ and then decreases. Similar
to $G_{N}$, the weighted QFI $G_{N}F_{Q,N}$ reaches maximum at $N\sim \bar{n}
$, and then decreases with the increase of $N$. As depicted in Fig.~\ref{fig2}%
(c), one can also see that the values of $G_{N}F_{Q,N}$ tend to vanishing as $%
N\gtrsim 5\bar{n}$, implying $N_{\mathrm{res}}\sim 5\bar{n}$.

To confirm the above result, we maximize Eq.~(\ref{QFIres}) with respect to $%
\alpha ^{2}$ for given $\bar{n}$ and $N_{\mathrm{res}}$. Figure~\ref{fig2}%
(d) shows $F_{Q}$ as a function of $\alpha ^{2}$ for a fixed $\bar{n}=5$,
where $N_{\mathrm{res}}=3\bar{n}$ (the solid circles), $5\bar{n}$ (the squares), and $10\bar{n}$ (the open circles). When $N_{\mathrm{res}}=\infty $
(the red solid line), the ideal result of the QFI is recovered and is given
by Eq.~(\ref{QFI}), which reaches the Heisenberg scaling $\alpha _{\mathrm{%
opt}}^{2}\simeq \bar{n}/2$~\cite{Smerzi08,Liu,Lang}. One can see that the
QFI with $N_{\mathrm{res}}=10\bar{n}$ almost follow the ideal result. In
Fig.~\ref{fig2}(e) and (f), we show optimal value of the ratio $\alpha^{2}/\bar{n}$ and the associated QFI $F_{Q,\mathrm{opt}}=F_{Q}(N_{\mathrm{%
res}},\bar{n},\alpha _{\mathrm{opt}}^{2})$ for each a given value of $\bar{n}%
\in \lbrack 1,10]$, where we take the number resolution $N_{\mathrm{res}}$ the same to Fig.~\ref%
{fig2}(d). From Fig.~\ref{fig2}(e), one can see that when $N_{\mathrm{res}}>%
\bar{n}$, the optimal input state contains more coherent light photons than
that of the squeezed vacuum. The Heisenberg scaling of the QFI is attainable
with $N_{\mathrm{res}}\gtrsim 5\bar{n}$, as depicted by Fig.~\ref{fig2}(f).

\begin{figure}[hptb]
\begin{centering}
\includegraphics[width=1\columnwidth]{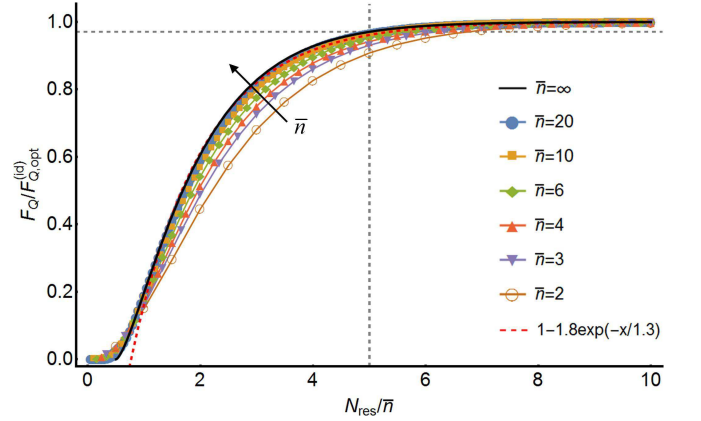}
\caption{Numerical results of $F_{Q}/F_{Q,\mathrm{opt}}^{\mathrm{(id)}}$ as a function of $N_{\mathrm{res}}/\bar{n}$ for given values of $\bar{n}$, using the optimal condition that maximizes Eq.~(\ref{QFI}). The solid line is given by our asymptotic result, i.e., Eq.~(\ref{ratio}), and the dashed line is a fitting result for the case $\bar{n}=20$. Both of them indicate that $96\%$ of the ideal QFI can be obtained as long as $x=N_{\mathrm{res}}/\bar{n}\gtrsim 5$ (the vertical dashed lines).} \label{fig3}
\end{centering}
\end{figure}

Figure~\ref{fig3} shows $F_{Q}/F_{Q,\mathrm{opt}}^{\mathrm{(id)}}$ as a
function of $N_{\mathrm{res}}/\bar{n}$ for the increase of $\bar{n}$ from $2$
up to $20$. For each a given $\bar{n}$, we first maximize the ideal QFI with
respect to $\alpha ^{2}$ to obtain $\alpha _{\mathrm{opt}}^{2}$ and $F_{Q,%
\mathrm{opt}}^{\mathrm{(id)}}$, as depicted by the red lines of Fig.~\ref%
{fig2}(d)-(f), and then calculate the QFI of Eq.~(\ref{QFIres}) using the
same input state. Our
numerical results show that $F_{Q}/F_{Q,\mathrm{opt}}^{\mathrm{(id)}}$
increases with $N_{\mathrm{res}}$ and approaches to $1$ as $N_{\mathrm{res}%
}\gg \bar{n}$.

To quantify how much phase information is kept by a finite cutoff $N_{%
\mathrm{res}}$, we try to find analytical result of $F_{Q}/F_{Q,\mathrm{opt}}^{%
\mathrm{(id)}}$ in the limit $\bar{n}\rightarrow \infty $. To this end, we
first separate the QFI into two terms $F_{Q}=F_{Q}^{\mathrm{(id)}}-F_{Q}^{%
\mathrm{(lost)}}$, where $F_{Q}^{\mathrm{(lost)}}=\sum_{N=N_{\mathrm{res}%
}}^{\infty }G_{N}F_{Q,N}$ denotes the QFI being lost. This expression is the same to Eq.~(\ref{QFIres}), except the sum over $N\in
(N_{\mathrm{res}},\infty )$. Next, we note that the photon number
distribution of the coherent state is much narrow than that of the squeezed
vacuum, which enable us to obtain an approximate result of $F_{Q}^{\mathrm{%
(lost)}}$ (see Appendix B). Furthermore, the ideal result of the QFI can
reach its maximum at the optimal condition $\alpha ^{2}=\sinh ^{2}\xi =\bar{n%
}/2\gg 1$~\cite{Smerzi08,Liu,Lang}. Using the same input, we obtain%
\begin{eqnarray}
\frac{F_{Q}}{F_{Q,\mathrm{opt}}^{\mathrm{(id)}}} &\approx&1-\lim_{\bar{n}%
\rightarrow \infty }\frac{F_{Q}^{\mathrm{(lost)}}(x,\bar{n})}{\bar{n}(\bar{n}%
+3/2)}  \notag \\
&\approx &\mathrm{erf}\left( \sqrt{x-1/2}\right) -\frac{2e^{-x+1/2}}{\sqrt{%
\pi }}\sqrt{x-1/2},  \label{ratio}
\end{eqnarray}%
where $x\equiv N_{\mathrm{res}}/\bar{n}$ and $\mathrm{erf}(...)$ denotes the
error function. Our analytical result shows a good agreement with the numerical results; see the solid lines of Fig.~\ref{fig3}. When $N_{\mathrm{res}}\gtrsim 5\bar{n}$, it predicts that over $96\%$ of
the ideal QFI can be kept; while for $N_{\mathrm{res}}<\bar{n}/2$, most of the phase information is lost.

Finally, it should be mentioned that coherent-state interferometry has been
demonstrated using two visible light photon counters with $N_{\mathrm{res}%
}=8 $~\cite{Smerzi07}. This number resolution is large enough to realize the
global phase estimation for the coherent-state input $\bar{n}\simeq 1$.
Based upon a Bayesian protocol~\cite{Smerzi07}, the achievable phase
sensitivity was found almost saturating quantum Cram\'{e}r-Rao bound over a
wide phase interval, in agreement with the theoretical prediction $F(\varphi
)=F_{Q}=\bar{n}$. To realize higher-precision optical metrology, it requires
a bright nonclassical light source with larger mean photon number~\cite%
{Smerzi08}, low photon loss~\cite{Donner,Joo,Zhang,Knott} and low noise~\cite%
{Qasimi,Teklu,YCLiu,Brivio,Genoni,Genoni12,Escher,Zhong,Bardhan,Feng2014,Vidrighin,YGao}%
, as well as the photon
counters with high detection efficiency~\cite{Calkins} and large enough number resolution.

\section{Conclusion}

In summary, we have investigated optical phase estimation with coherent $%
\otimes $ squeezed vacuum light by using imperfect photon counters with an
upper threshold $N_{\mathrm{res}}$ for the photon number resolution. We show
that both the CFI and the QFI are the sum of the contributions from
individual $N$-photon components, and the CFI always saturates the QFI for
each individual $N$-photon component. For ideal photon-counting detectors with $N_{%
\mathrm{res}}\rightarrow \infty $, the CFI or the QFI attains its maximum $%
F_{Q,\mathrm{opt}}^{\mathrm{(id)}}\sim \bar{n}^{2}$ when $|\alpha |^{2}\simeq \sinh ^{2}|\xi |$, leading to the
Heisenberg limit of the estimation precision. For the detectors with large enough number resolution $N_{\mathrm{res}}>\bar{n}$, we find that the optimal input state
contains more coherent light photons than that of the squeezed vacuum. We present an analytical result that quantifies the amount of lost information and show that over $96$ percent of ideal QFI can be attained as long as $N_{\mathrm{res}}\gtrsim 5\bar{n}$; While for $N_{\mathrm{res}}<\bar{n}/2$, most of the phase information is lost. Our results highlight the important influence of
the finite number resolution of photon-counting detectors on optical phase estimation. It is also interesting to explore the
performance of other continuous-variable input states, e.g., a product of
two squeezed vacuum $|\xi \rangle \otimes |-\xi \rangle $~\cite{Lang2}, when
realistic photon counters are used.

\begin{acknowledgments}
The first two authors P. Liu and P. Wang contribute equally to this work. We
would like to thank S. Rosen and Y. Silberberg for kindly response to our
questions. This work has been supported by the NSFC (Grant Nos. 11421063,
11534002, 11274036, and 11322542), the National 973 program (Grant Nos.
2012CB922104, 2014CB921403, and 2014CB848700). G.R.J. also acknowledges
support from the Major Research Plan of the NSFC (Grant No. 91636108).
\end{acknowledgments}

\appendix

\begin{widetext}

\section{The $N$-photon state under the phase-matching condition}

Formally, a single-mode squeezed vacuum of light is defined by $|\xi \rangle
=S(\xi )|0\rangle $, with the squeeze operator~\cite{RAFisher,Truax,Ban}:
\begin{equation}
S(\xi )=\exp \left[ \frac{1}{2}(\xi ^{\ast }b^{2}-\xi b^{\dag 2})\right]
=\exp \left( -e^{i\theta _{b}}\frac{\tanh |\xi |}{2}b^{\dagger 2}\right)
\left( \frac{1}{\cosh |\xi |}\right) ^{b^{\dagger }b+\frac{1}{2}}\exp \left(
e^{-i\theta _{b}}\frac{\tanh |\xi |}{2}b^{2}\right) ,  \label{disentangling}
\end{equation}%
where $\xi =|\xi |\exp (i\theta _{b})$ denotes complex amplitude of the
squeezed vacuum. In Fock basis, using $b|0\rangle =0$, the squeezed vacuum
can be expressed as
\begin{equation}
|\xi \rangle =\frac{1}{\sqrt{\cosh |\xi |}}\exp \left( -e^{i\theta _{b}}%
\frac{\tanh |\xi |}{2}b^{\dagger 2}\right) |0\rangle =\sum_{k=0}^{+\infty
}s_{2k}|2k\rangle ,
\end{equation}%
where $s_{2k}\equiv \langle 2k|\xi \rangle $ denote the probability
amplitudes of the squeezed vacuum, given by
\begin{equation}
s_{2k}(\theta _{b})=\frac{\sqrt{(2k)!}}{k!\sqrt{\cosh |\xi |}}\left(
-e^{i\theta _{b}}\frac{\tanh |\xi |}{2}\right) ^{k},\hskip12pt\text{or}\hskip%
6pts_{k}(\theta _{b})=\frac{H_{k}(0)}{\sqrt{k!\cosh |\xi |}}\left(
e^{i\theta _{b}}\frac{\tanh |\xi |}{2}\right) ^{k/2},  \label{ss}
\end{equation}%
with the Hermite polynomials $H_{2n}(0)=(-1)^{n}(2n)!/n!$ and $H_{2n+1}(0)=0$%
.

Note that one can obtain explicit form of the squeezed vacuum using the
disentangling formula~\cite{RAFisher,Truax,Ban}, as done in Eq.~(\ref%
{disentangling}), or alternatively, directly solving the eigenvalue equation
$S(\xi )b|0\rangle =S(\xi )bS^{\dagger }(\xi )|\xi \rangle =0$~\cite{Gerry}.
The single-mode squeezed vacuum contains only even number of photons and has
been generated in experiments~\cite%
{Slusher1985,LAWu1986,LAWu1987,Slusher1987,Breitenbach97,Vahlbruch08,Vahlbruch16}%
.

We now consider the interferometer fed with the squeezed vacuum from one
input port and a coherent-state light from another port. The coherent state
is given by $|\alpha \rangle =\sum_{n}c_{n}(\theta _{a})|n\rangle $, with
the probability amplitudes%
\begin{equation}
c_{n}(\theta _{a})\equiv \langle n|\alpha \rangle =e^{-\left\vert \alpha
\right\vert ^{2}/2} \frac{|\alpha |^{n}e^{in\theta _{a}}}{\sqrt{n!}},
\label{cs}
\end{equation}%
where $\alpha =|\alpha |\exp (i\theta _{a})$ denotes the complex amplitude
of the coherent light. In Eqs.~(\ref{ss}) and ~(\ref{cs}), we have written
down explicitly the phase dependence of the probability amplitudes, purely
for later use.

Under the phase-matching condition (PMC): $\cos (\theta _{b}-2\theta
_{a})=+1 $, we now calculate the probability amplitudes of the $N$-photon
states $|\psi _{N}\rangle $ as
\begin{eqnarray}
\frac{c_{N-2k}(\theta _{a})s_{2k}(\theta _{b})}{\sqrt{G_{N}}} &=&(-1)^{k}%
\frac{e^{iN\theta _{a}}e^{ik(\theta _{b}-2\theta _{a})}}{\sqrt{R_{N}(x)}}%
\frac{\sqrt{(2k)!}}{k!\sqrt{(N-2k)!}}\left( 2x\right) ^{(N-2k)/2}  \notag \\
&\overset{\text{\textrm{PMC}}}{\longrightarrow }&(-1)^{k}\frac{e^{iN\theta
_{a}}}{\sqrt{R_{N}(x)}}\frac{\sqrt{(2k)!}}{k!\sqrt{(N-2k)!}}\left( 2x\right)
^{(N-2k)/2} \equiv e^{iN\theta _{a}}\frac{c_{N-2k}(0)s_{2k}(0)}{\sqrt{G_{N}}}%
,  \label{CSG}
\end{eqnarray}%
where we have used explicit form of $G_{N}$, given by Eq.~(\ref{GN}), and
the condition $\exp [ik(\theta _{b}-2\theta _{a})]=+1$ for integers $k$.
Note that Eq.~(\ref{psi_N}) can be rewritten as $|\psi _{N}\rangle =\exp
(iN\theta _{a})|\tilde{\psi}_{N}\rangle $, where $|\tilde{\psi}_{N}\rangle $
denotes the $N$-photon states with real amplitudes (i.e., $\theta
_{a}=\theta _{b}=0$).

Finally, we consider a unitary operation $\exp (-i\varphi J_{\eta })$ on the
$N$-photon states $|\psi _{N}\rangle $, with $J_{\eta }=J_{x}\cos \eta
+J_{y}\sin \eta $, to obtain Eqs.~(\ref{ampBS}) and (\ref{pmu}) in main
text. Under the PMC, we obtain
\begin{eqnarray}
e^{-i\varphi J_{\eta }}|\psi _{N}\rangle &=&e^{iN\theta _{a}}e^{-i\varphi
J_{\eta }}|\tilde{\psi}_{N}\rangle =e^{iN\theta _{a}}e^{-i\eta
J_{z}}e^{-i\varphi J_{x}}e^{i\eta J_{z}}|\tilde{\psi}_{N}\rangle  \notag \\
&=&\frac{e^{iN\theta _{a}}}{\sqrt{G_{N}}}e^{-i\eta J_{z}}e^{-i\varphi
J_{x}}\sum_{k=0}^{[N/2]}e^{i\eta (J-2k)}c_{N-2k}(0)s_{2k}(0)|J,J-2k\rangle ,
\end{eqnarray}%
where, in the second step, we have used the relation $\exp (-i\eta
J_{z})f(J_{x})\exp (i\eta J_{z})=f(J_{\eta })$, and Eq.~(\ref{psi_N}) with $%
\theta _{a}=\theta _{b}=0$ for $|\tilde{\psi}_{N}\rangle $, which is
expressed in terms of the states $|J,J-2k\rangle =|N-2k\rangle _{a}\otimes
|2k\rangle _{b}$. In the eigenbasis of $J_{z}$, we obtain the probability
amplitudes
\begin{eqnarray}
\langle J,\mu |e^{-i\varphi J_{\eta }}|\psi _{N}\rangle &=&\frac{e^{iN\theta
_{a}}}{\sqrt{G_{N}}}e^{-i\eta \mu }\sum_{k=0}^{[N/2]}e^{i\eta
(J-2k)}c_{N-2k}(0)s_{2k}(0)\langle J,\mu |e^{-i\varphi J_{x}}|J,J-2k\rangle
\notag \\
&=&\frac{e^{iN\theta _{a}}}{\sqrt{R_{N}(x)}}e^{i(\frac{\pi }{2}-\eta )(\mu
-J)}\sum_{k=0}^{[N/2]}e^{-i2k\eta }\frac{\sqrt{(2k)!}}{k!\sqrt{(N-2k)!}}%
\left( 2x\right) ^{N/2-k}d_{\mu ,J-2k}^{J}(\varphi ),  \label{amplitude}
\end{eqnarray}%
where, in the last step, we have introduced Wigner's d-matrix $d_{\mu
,v}^{J}(\varphi )$. Obviously, for the special case $\eta =0$ and $\varphi
=\pi /2$, we obtain the $N$-photon state after the first 50:50 beam splitter
$\exp (-i\pi J_{x}/2)|\psi _{N}\rangle $ and its probability distributions;
see Eqs.~(\ref{ampBS}) and (\ref{pmu}). For $\eta =\pi /2$ and arbitrary $%
\varphi $, we can obtain the output state $\exp (-i\varphi J_{y})|\psi
_{N}\rangle $ and its probabilities $P_{N}(\mu |\varphi )$.

\section{Analytical results of the quantum Fisher information}

In a lossless and noiseless interferometer, the QFI of a pure phase-encoded
state $|\psi _{\mathrm{out}}\rangle =\exp (-i\varphi G)|\psi _{\mathrm{in}%
}\rangle $ is simply given by $F_{Q}=4(\langle G^{2}\rangle _{\mathrm{in}%
}-\langle G\rangle _{\mathrm{in}}^{2})$~\cite%
{Braunstein,Braunstein96,Luo,Smerzi09,Donner}, where $G$ is a Hermitian
operator. For the squeezed-state interferometer, as illustrated in Fig.~\ref%
{fig1}(a), the input state is the product of a coherent state and a squeezed
vacuum, i.e., $|\psi _{\mathrm{in}}\rangle =|\alpha \rangle _{a}\otimes |\xi
\rangle _{b}$, and the phase shifter is given by $G=J_{x}$, or $J_{y}$,
where, for brevity, we have introduced the angular-momentum operators $%
J_{+}=(J_{-})^{\dag }=a^{\dag }b$ and $J_{z}=(a^{\dag }a-b^{\dagger }b)/2$,
with the bosonic operators of two light fields $a$ and $b$.

According to Ref.~\cite{Smerzi08}, the QFI of the output state $\exp
(-i\varphi J_{y})|\psi _{\mathrm{in}}\rangle $ is optimal when the two
injected light fields are phase matched, i.e., the PMC $\cos (\theta
_{b}-2\theta _{a})=+1$. Recently, Liu \textit{et al}.~\cite{Liu} have
derived a more general form of the PMC for the interferometer $U_{\mathrm{MZI%
}}(\varphi )=\exp (-i\varphi J_{y})$, where a superposition of even or odd
number of photons is injected from one port and an arbitrary state from
another port.

To show it clearly, we focus on the PMC $\cos (\theta _{b}-2\theta _{a})=+1$
and calculate the QFI of $\exp (-i\varphi J_{y})|\psi _{\mathrm{in}}\rangle $%
, namely
\begin{equation}
F_{Q}^{\mathrm{(id)}}=4\langle J_{y}^{2}\rangle _{\mathrm{in}}=\langle
(a^{\dag }a+b^{\dag }b+2a^{\dag }ab^{\dag }b)-(a^{\dag 2}b^{2}+H.c.)\rangle
_{\mathrm{in}},
\end{equation}%
where $H.c.$ denotes the Hermitian conjugate. There are two contributions to
the QFI. First, it is easy to obtain
\begin{equation}
\langle (a^{\dag }a+b^{\dag }b+2a^{\dag }ab^{\dag }b)\rangle _{\mathrm{in}}=%
\bar{n}_{a}+\bar{n}_{b}+2\bar{n}_{a}\bar{n}_{b},
\end{equation}%
with $\bar{n}_{a}=|\alpha |^{2}$ and $\bar{n}_{b}=\sinh ^{2}|\xi |$ being
mean photon number of light fields from two input ports. Second, using the
relation $S^{\dag }(\xi )bS(\xi )=b\cosh |\xi |-b^{\dag }e^{i\theta
_{b}}\sinh |\xi |$, we obtain
\begin{equation}
\langle a^{\dag 2}b^{2}\rangle _{\mathrm{in}}=\alpha ^{\ast 2}\left\langle
\xi \right\vert b^{2}\left\vert \xi \right\rangle =-\bar{n}_{a}\sqrt{\bar{n}%
_{b}(1+\bar{n}_{b})}e^{i(\theta _{b}-2\theta _{a})}.  \label{a2b2}
\end{equation}%
Therefore, the ideal result of the QFI is given by%
\begin{equation}
F_{Q}^{\mathrm{(id)}}=\bar{n}_{a}\left[ 1+2\bar{n}_{b}+2\sqrt{\bar{n}_{b}(1+%
\bar{n}_{b})}\cos \left( \theta _{b}-2\theta _{a}\right) \right] +\bar{n}%
_{b}\leq \bar{n}_{a}\left[ 1+2\bar{n}_{b}+2\sqrt{\bar{n}_{b}(1+\bar{n}_{b})}%
\right] +\bar{n}_{b},  \label{idealQFI}
\end{equation}%
where the equality holds when the PMC is fulfilled, i.e., $\cos (\theta
_{b}-2\theta _{a})=+1$. Similarly, one can note that the PMC $\cos (\theta
_{b}-2\theta _{a})=-1$ is a good choice for the output state $\exp
(-i\varphi J_{x})|\alpha \rangle _{a}\otimes |\xi \rangle _{b}$, e.g., the
phases of the two light fields $(\theta _{a}, \theta _{b})=(0, \pi)$~\cite{Hofmann} and $(\pi
/2, 0)$~\cite{Afek}. Furthermore, one can simplify the ideal result of the
QFI as Eq.~(\ref{QFI}), using the relation $1+2\bar{n}_{b}+2\sqrt{\bar{n}%
_{b}(1+\bar{n}_{b})}=e^{2|\xi |}$.

With a finite number resolution $N_{\mathrm{res}}$, we have shown that the
CFI and the QFI are the same and is given by Eqs.~(\ref{QFIres}), which can
be rewritten as%
\begin{eqnarray}
F_{Q} &=&\sum_{N_{a}=0}^{N_{\mathrm{res}}}\sum_{N_{b}=0}^{N_{\mathrm{res}}-N_{a}}%
\left[ N_{a}+\left( 1+2N_{a}+\frac{2\alpha ^{2}}{\tanh \xi }\right) N_{b}%
\right] \left[ c_{N_{a}}(0)s_{N_{b}}(0)\right] ^{2}  \notag \\
&\approx &\bar{n}_{a}\sum_{N_{b}=0}^{N_{\mathrm{res}}-\bar{n}_{a}}\left[
s_{N_{b}}(0)\right] ^{2}+\left( 1+2\bar{n}_{a}+\frac{2\bar{n}_{a}}{\tanh \xi
}\right) \sum_{N_{b}=0}^{N_{\mathrm{res}}-\bar{n}_{a}}N_{b}\left[
s_{N_{b}}(0)\right] ^{2},  \label{QFI2}
\end{eqnarray}%
where, for brevity, we consider the two light fields with real amplitudes, i.e., $\theta
_{b}=\theta _{a}=0$, and the probability amplitudes $c_{n}(0)$ and $s_{k}(0)$
are given by Eqs.~(\ref{ss}) and~(\ref{cs}). In the above result, we made
an approximation
\begin{equation}
\sum_{N_{a}=0}^{N_{\mathrm{res}}}\sum_{N_{b}=0}^{N_{\mathrm{res}%
}-N_{a}}f(N_{a})g(N_{b})\left[ c_{N_{a}}(0)s_{N_{b}}(0)\right] ^{2}\approx\sum_{N_{a}=0}^{\infty}f(N_{a})\left[ c_{N_{a}}(0)\right] ^{2} \sum_{N_{b}=0}^{N_{\mathrm{res}}-\bar{n}_{a}}g(N_{b})\left[s_{N_{b}}(0)\right] ^{2},
\end{equation}%
where $\bar{n}_{a}=|\alpha |^{2}$ and the sum over the mode $b$ is still kept, since the photon number distribution of the squeezed vacuum is usually wider than
that of the coherent state (even for $\bar{n}_b<\bar{n}_a$)~\cite{Gerry,Breitenbach97}. For a finite $\bar{n}_{a}$ and $%
N_{\mathrm{res}}\rightarrow \infty $, it is easy to obtain the ideal result
of the QFI as%
\begin{equation}
F_{Q}\approx \bar{n}_{a}+\left( 1+2\bar{n}_{a}+\frac{2\bar{n}_{a}}{\tanh \xi
}\right) \bar{n}_{b}=F_{Q}^{\mathrm{(id)}},  \label{back to id}
\end{equation}%
where $\tanh \xi =\sqrt{\bar{n}_{b}/(\bar{n}_{b}+1)}$ and $F_{Q}^{\mathrm{%
(id)}}$ is given by Eq.~(\ref{idealQFI}).

Finally, we consider a finite number resolution with large enough $N_{%
\mathrm{res}}$ ($>\bar{n}_{a}$), and derive analytical result of the QFI. To
this end, we first rewrite Eq.~(\ref{QFI2}) as $F_{Q}=F_{Q}^{\mathrm{(id)}%
}-F_{Q}^{\mathrm{(lost)}}$, where $F_{Q}^{\mathrm{(lost)}}$ quantifies the
lost phase information caused by the finite number resolution, given by
\begin{eqnarray}
F_{Q}^{\mathrm{(lost)}} &=&4\sum_{N=N_{\mathrm{res}}}^{\infty }G_{N}\langle
\psi _{N}|J_{y}^{2}|\psi _{N}\rangle   \notag \\
&\approx &\bar{n}_{a}\sum_{k=N_{\mathrm{res}}-\bar{n}_{a}}^{\infty }\left[
s_{k}(0)\right] ^{2}+\left( 1+2\bar{n}_{a}+\frac{2\bar{n}_{a}}{\tanh \xi }%
\right) \sum_{k=N_{\mathrm{res}}-\bar{n}_{a}}^{\infty }k\left[ s_{k}(0)%
\right] ^{2}  \notag \\
&\approx &\left( 2\bar{n}_{a}+\frac{2\bar{n}_{a}}{\tanh \xi }\right)
\int_{N_{\mathrm{res}}-\bar{n}_{a}}^{\infty }kdk\frac{\left( \tanh \xi
\right) ^{k}}{\sqrt{2\pi k}\cosh \xi },  \label{QFIlost0}
\end{eqnarray}%
where, in the last step, we only keep the terms $\sim O(\bar{n}^{2})$. In
addition, we replace the sum over $k$ by an integral and use the Stirling's
formula $k!\approx \sqrt{2k\pi }(k/e)^{k}$. When $N_{\mathrm{res}}\leq \bar{n%
}_{a}$, it is easy to find $F_{Q}^{\mathrm{(lost)}}\approx F_{Q}^{%
\mathrm{(id)}}$ and hence the achievable QFI $F_{Q}\sim O(\bar{n}^{0})$ or $O(\bar{n}^{1})$, corresponding to almost complete loss of the phase information, or the ultimate estimation precision in the classical limit. To enlarge the QFI, we take $N_{\mathrm{res}}>\bar{n}_{a}$
and obtain
\begin{equation}
F_{Q}^{\mathrm{(lost)}}\approx \frac{2\bar{n}_{a}\bar{n}_{b}}{B^{3/2}}\left(
1+e^{-\frac{B}{2\bar{n}_{b}}}\right) \left[ \mathrm{erfc}\left( \sqrt{\frac{%
N_{\mathrm{res}}-\bar{n}_{a}}{2\bar{n}_{b}}B}\right) +\frac{2}{\sqrt{\pi}}%
e^{-\frac{N_{\mathrm{res}}-\bar{n}_{a}}{2\bar{n}_{b}}B}\sqrt{\frac{N_{%
\mathrm{res}}-\bar{n}_{a}}{2\bar{n}_{b}}B}\right] ,  \label{QFIlost1}
\end{equation}%
where $\bar{n}_{b}=\sinh ^{2}\xi $, $B(\bar{n}_{b})=\bar{n}_{b}\log [(1+\bar{n}_{b}) /\bar{n}_{b}]$, and $\mathrm{erfc}(x)=1-\mathrm{erf}(x)$ denotes the
complementary error function. Our analytical result coincides with the numerical results in Figs.~\ref{fig2}(d)-(f). In the limit $\bar{n}_{a}=\bar{n}_{b}=\bar{n}/2\rightarrow\infty$, we obtain $B(\bar{n}_{b})\rightarrow 1$ and hence
\begin{equation}
F_{Q}^{\mathrm{(lost)}}\approx \bar{n}^{2}\left[ \mathrm{erfc}\left( \sqrt{%
x-1/2}\right) +\frac{2e^{-(x-1/2)}}{\sqrt{\pi }}\sqrt{(x-1/2)}\right] ,
\end{equation}%
where $x\equiv N_{\mathrm{res}}/\bar{n}>1/2$. This result gives Eq.~(\ref%
{ratio}) in main text.

\end{widetext}

\end{document}